\documentclass[twocolumn,showpacs,amsmath,superscriptaddress,
amssymb,nofootinbib]{revtex4-1}
\usepackage{dcolumn}
\usepackage{bm}
\usepackage{graphicx}

\usepackage{color}

\newcommand{\be}{\begin{equation}}
\newcommand{\ee}{\end{equation}}
\newcommand{\ba}{\begin{eqnarray}}
\newcommand{\ea}{\end{eqnarray}}
\newcommand{\hh}{\, ,\hspace{0.5cm}}
\newcommand{\hhh}{\, ,\hspace{0.2cm}}
\newcommand{\pa}{\partial}

\newcommand{\n}[1]{\label{#1}}

\newcommand{\CAL}{\mathcal}

\newcommand{\inds}[1]{{\scriptscriptstyle #1}}
\newcommand{\lap}{\bigtriangleup}
\newcommand{\lan}{\langle}
\newcommand{\ran}{\rangle}
\newcommand{\llan}{\langle\!\langle}
\newcommand{\rran}{\rangle\!\rangle}

\begin{document}

\title{Mass-gap for black hole formation in higher derivative and ghost free gravity}
\author{Valeri P. Frolov}
\affiliation{Department of Physics, University of Alberta, Edmonton,
Alberta, Canada T6G 2E1}
\email{vfrolov@ualberta.ca}


\begin{abstract}
We study a spherical gravitational collapse of a small mass in higher derivative and ghost free theories of gravity. By boosting a solution of linearized equations for a static point mass in such theories we obtain in the Penrose limit the gravitational field of an ultra-relativistic particle. Taking a superposition of such solutions we construct a metric of a collapsing null shell in the linearized higher derivative and ghost free gravity. The latter allows one to find the gravitational field of a thick null shell. By analysing these solutions we demonstrate that in a wide class of the higher dimensional theories of gravity as well as for the ghost free gravity there exists a mass gap for the mini black hole production. We also found conditions when the curvature invariants remain finite at $r=0$ for the collapse of the thick null shell.
\end{abstract}

\pacs{04.70.s, 04.70.Bw, 04.20.Jb \hfill  Alberta-Thy-7-15}

\maketitle

It is generally believed that the theory of general relativity (GR) should be modified to improve its ultraviolet (UV) behavior and remove singularities.
One of the options is to allow terms in the gravitational action that contain more than two derivations. The UV properties of the higher derivative theory of gravity are usually better than in GR. In particular, the forth order gravity can be made renormalizable \cite{ST_1}. At the same time, the gravitational potential of a point mass in the Newtonian limit of such theories is usually finite (see e.g. \cite{ST_2,SHAP} and references therein). However the higher derivative gravity possesses new unphysical degrees of freedom (ghosts) \cite{ST_1,ST_2}. The problem of ghosts can be solved if one allows an infinite number of derivatives in the gravity action, that makes it non-local. Ghost-free theories of gravity are discussed in \cite{Tomb,GFG_1,BKM,GFG_2}. Their application to the problem of singularities in cosmology and black holes can be found in \cite{SING}

In this paper we study gravitational collapse of a small mass in higher derivative (HD) and host-free (GF) theories of gravity. We obtain solutions of the linearized equations for such theories for a spherical collapse of null fluid.  We demonstrate that if a static gravitational field of a point mass in the HD and GF gravity is regular at $r=0$ \cite{SHAP}, then the metric for the collapsing object has the same property. This means, that the perturbation of the metric, which is proportional to the collapsing mass $M$, is smooth and uniformly bounded, so that the higher in $M$ corrections can be neglected in the leading order. This implies that for the collapse of a small mass an apparent horizon is not formed. In other words, for this wide class of HD and GF theories of gravity there exists a mass gap for mini black hole production. This property is a consequence of the existence of the UV length scale, where such theories become different from GR. For the Weyl modified gravity this was shown long time ago in \cite{FV}.

We study the linearized gravity equations on the flat Minkowski background $\eta_{\mu\nu}$ and write the metric in the form $g_{\mu\nu}=\eta_{\mu\nu}+h_{\mu\nu}$.
The most general action for the higher derivative theory of gravity which contains not higher than second power of $h_{\mu\nu}$ is \cite{GFG_1,BKM}
\ba
S=&-&\int d^4x \left[{1\over 2}h_{\mu\nu} a(\Box)\Box h^{\mu\nu}+h_{\mu}^{\, \sigma} b(\Box)\pa_{\sigma}\pa_{\nu}h^{\mu\nu} \right.\nonumber\\
&+&h c(\Box)\pa_{\mu}\pa_{\nu}h^{\mu\nu}+
{1\over 2} h d(\Box)\Box h\nonumber\\
&+&\left.
h^{\lambda\sigma}{f(\Box)\over \Box}\pa_{\sigma}\pa_{\lambda}\pa_{\mu}\pa_{\nu}h^{\mu\nu}\right]\, .\n{SSQ}
\ea
Five, in general non-linear functions of the box-operator obey the following three relations
\be\n{abc}
a+b=0\hhh c+d=0\hhh b+c+f=0\, .
\ee
Thus the action $S$ contains in fact only two independent arbitrary functions of the box operator. In order to recover GR in the infra-red domain these functions must satisfy the following conditions $a(0)=c(0)=-b(0)=-d(0)=1$. Let us list some special interesting examples \cite{BKM}:
\begin{enumerate}
\item General relativity (GR), ${\CAL L}=R$: $a=c=1$;
\item Gauss-Bonnet (GB) gravity, ${\CAL L}=R+\alpha(\Box){\CAL G}$, where ${\CAL G}=R^2-4R_{\mu\nu}R^{\mu\nu}+R_{\mu\nu\alpha\beta}R^{\mu\nu\alpha\beta}$ is the Gauss-Bonnet invariant: $a=c=1$;
\item $L(R)$ gravity, ${\CAL L}(R)={\CAL L}(0)+{\CAL L}'(0)R+1/2{\CAL L}''(0) R^2+\ldots$: $a=1$, $c=1-{\CAL L}''(\Box)$;
\item Weyl gravity, ${\CAL L}=R-\mu^{-2}C_{\mu\nu\alpha\beta}C^{\mu\nu\alpha\beta}$: $a=1-\mu^{-2}\Box$, $c=1-{1\over 3}\mu^{-2}\Box$;
\item Higher derivative (HD) gravity: $a=\prod_{i=1}^{n} (1-\mu_i^{-2}\Box)$,
$c=\prod_{k=1}^{n_c} (1-\nu_k^{-2}\Box)$. For simplicity, in what follows we assume that masses $\mu_i$ are different;
\item Ghost free gravity: $a=c=\exp(-\Box/\mu^2)$.
\end{enumerate}
It is evident that the  linearized GB gravity has the same properties as GR. In the same linearized approximation the Weyl and $L(R)$ theories of gravity are nothing but special cases of the general HD gravity.

Let us consider first static solutions of the linearized gravity equations. In the Newtonian limit the stress-energy tensor is $\tau_{\mu\nu}=\rho(\vec{r})\delta_{\mu}^0\delta_{\nu}^0$, and the metric is of the form
\be\n{dss}
ds^2=-(1+2\varphi) dt^2+(1-2\psi+2\varphi)d\ell^2\, .
\ee
The functions $\varphi$ and $\psi$ obey the equations \cite{GFG_1}\footnote{
Notice that the first of the equations (17) of this paper contains a misprint and it should be written as
$2(a-3c)[\nabla^2\Phi-2\nabla^2\Psi]=\kappa\rho$.
}
\ba
&&a(\lap)\lap \psi=8\pi G \rho\, ,\n{mast}\\
&&(a(\lap)-3c(\lap))(\lap\varphi-2\lap\psi)=8\pi G\rho\, .\n{sec}
\ea
Here $\lap$ is a usual flat Laplace operator in a flat 3D space with metric $d\ell^2$,  and $G$ is the gravitational coupling constant.
After solving (\ref{mast}) and finding the potential $\psi$, one can find the second potential $\varphi$ by solving (\ref{sec}).

For a point mass $\rho=m \delta(\vec{r})$ the solution (\ref{dss}) is spherically symmetric. We call it finite if $\varphi(r)$ and $\psi(r)$ near $r=0$ have the form
\ba
&&\psi(r)\sim \psi_0+\psi_1 r+{1\over 2}\psi_2 r^2+O(r^3)\, ,\n{psir}\\
&&\varphi(r)\sim \varphi_0+\varphi_1 r+{1\over 2}\varphi_2 r^2+O(r^3)\, .\n{pvp}
\ea
A finite solution is not necessary regular one. Really, the Kretschmann tensor invariant  ${\CAL R}^2=R_{\alpha\beta\gamma\delta}R^{\alpha\beta\gamma\delta}$ for the metric
(\ref{dss}), (\ref{psir}) and (\ref{pvp}) is the form
\ba
&&{\CAL R}^2={A_2\over r^2}+{A_1\over r}+O(1)\, ,\n{AAA}\\
&&A_2=8(4\psi_1^2-5\psi_1\varphi_1+3\varphi_1^2)\, ,\nonumber\\
&&A_1=16[\psi_1(5\psi_2-4\varphi_2)-4\varphi_1(\psi_2 -\varphi_2)]\nonumber\, .
\ea
The quantity $A_2$ is a positive definite quadratic form of variables $\psi_1$ and $\varphi_1$, and it vanishes only when $\psi_1=\varphi_1=0$. In such a case the quantity $A_1$ vanishes as well, so that ${\CAL R}^2$ is finite at $r=0$. We call such a solution regular. We also call a solution $\psi$-regular, if $\psi_1=0$. For a special class of theories, where $a=c$, one has $\psi=2\varphi$ and a solution which is $\psi$-regular is at the same time a regular one.

We denote $\hat{O}=a(\lap)\lap$,
\be\n{QQQ}
Q(\xi)=\hat{O}^{-1}(\lap=-\xi)=-[\xi a(-\xi)]^{-1}\, .
\ee
and assume that $Q(\xi)$ can be written as the Laplace transform of some function $f(s)$
\ba\n{QQf}
&&Q(\xi)=\int_0^{\infty} ds f(s) e^{-s\xi}\, ,\\
&&f(s)={1\over 2\pi i}\int_{\alpha-i\infty}^{\alpha+i\infty}d\xi Q(\xi) e^{s\xi}\, .\n{INV}
\ea
The second relation is nothing but the inverse Laplace transform. A parameter $\alpha$ must be chosen so that the integration path in (\ref{INV}) lies in the domain of the analyticity of $Q(\xi)$.

A formal solution of the operator equation
$\hat{O}\hat{G}=-\hat{I}$ can be written by using the Laplace transform (\ref{QQf}). It contains the exponent $\exp(s\lap)$, which in the $x$-representation is nothing but the heat kernel
\be
\lan x'|e^{s\lap}|x\ran=K(|x-x'|;s)={e^{-|x-x'|^2/(4s)}\over (4\pi s)^{3/2}}\, \, .
\ee
Thus the potential $\psi(r)$ for a point mass is
\ba
\psi(r)&=&8\pi G m \int_0^{\infty} ds f(s) K(r;s)\n{fKK}\\
&=&{Gm\over \pi i r}\int_{\alpha-i\infty}^{\alpha+i\infty}d\xi Q(\xi) e^{-\sqrt{-\xi}r}\, .\n{pps}
\ea

We consider at first a case of HD gravity. We assume that the function $Q(\xi)$ has simple poles and write it in the form
\be\n{Qnn}
Q(\xi)=-[\xi \prod_{i=1}^n (1+\xi/\mu_i^2)]^{-1}\, .
\ee
This covers the above listed gravitational theories  1-5 except for some degenerate cases.

The Heaviside expansion theorem \cite{POUL} gives the following expression for $f(s)$
\be\n{ffss}
f(s)=-(1-\sum_{i=1}^n P_i^{-1} e^{-\mu_i t})\, ,
\ee
where $P_i=\prod_{j=1, j\ne i}^n (1-\mu_j^2/\mu_i^2)$. Taking the integral (\ref{fKK}) one obtains
\be\n{psps}
\psi(r)=-2Gm r^{-1}(1-\sum_{i=1}^n P_i^{-1} e^{-\mu_i r})\, ,
\ee

For GR $f(s)=1$ and one has
\be
\psi(r)=2\varphi(r)=-2Gm/r\, .
\ee
For a theory with higher derivatives, where $n\ge 1$, the potential $\psi(r)$ near $r=0$ has a form (\ref{psir}) with
\be
\psi_0= -2Gm S_1\, ,\ \
\psi_1=Gm S_2\, ,\ \
S_k=\sum_{i=1}^n \mu_i^k P_i^{-1}\, .
\ee
We used here a relation $S_0=1$. Thus a theory with higher derivatives is $\psi$-regular, when the  condition $S_2=0$ is satisfied.

For the GF gravity  $f(s)=-\vartheta(s-\mu^{-2})$ and one reproduces the result of \cite{GFG_1}
\be
\psi(r)=2\varphi(r)=-2Gm\ \mbox{erf}(\mu r/2)/r\, .
\ee
This solution is regular at $r=0$.

We demonstrate now, how using a solution of (\ref{mast}) for a static point mass one can obtain a solution for an ultra-relativistic particle.
Let us write the flat space metric in the form $d\ell^2=dy^2+d\zeta_{\inds{\perp}}^2$, and suppose that the source, generating the gravitational field (\ref{dss}), moves along the $y$-axis with a constant velocity $\beta$. To find the gravitational field of the moving source we make the following boost transformation
\be
t=\lambda_- v+\lambda_+u\hhh y=\lambda_- v-\lambda_+u\, .
\ee
Here $\lambda_{\pm}=(1\pm \beta)\gamma/2$ and $\gamma=(1-\beta^2)^{-1/2}$. In the limit $\gamma\to \infty$ one gets $y\sim -\gamma u$, $t\sim \gamma u$, $\ell^2\sim \gamma^2 u^2+\zeta_{\inds{\perp}}^2$, and
\ba\n{METR}
&&ds^2=-du dv +d\zeta_{\inds{\perp}}^2+dh^2\, ,\nonumber\\
&&dh^2=\Phi du^2\, ,\ \
\Phi=-2\lim_{\gamma\to \infty}(\gamma^2 \psi)\, .\n{PPHI}
\ea
We assume that the energy of the particle, $M=\gamma m$ remains constant in this (Penrose) limit. We use also the following relation
\be
\lim_{\gamma\to\infty} \gamma \exp(-\gamma^2 u^2/(4s))=\sqrt{4\pi s}\delta(u)\, .
\ee
These relations and (\ref{fKK}) give
\ba
&&\Phi=-4 G M F(\zeta_{\inds{\perp}}^2) \delta(u)\, ,\n{PP}\\
&&F(z)=\int_0^{\infty} {ds\over s} f(s)\, e^{-z/(4s)}\, .\n{PPP}
\ea
For GR (as well as for GB and $L(R)$ gravity) one has $f(s)=1$ and $F(z)=\ln(z/\eta^2)$, where $\eta$ is an infrared cut-off parameter. The relations (\ref{PPHI}) and (\ref{PP})  correctly reproduce the well known Aichelburg-Sexl solution for the gravitational field of an ultra-relativistic particle ("photon") in GR.

Using expression (\ref{ffss}) for $f(s)$ for the HD gravity and taking integral in (\ref{PPP}) one finds
\be\n{FHDG}
F(z)=\ln(z/\eta^2)+2\sum_{i=1}^n P_i^{-1}K_0(\mu_i \sqrt{z})\, .
\ee
In the limit $\mu_i\to\infty$ the second term in the right-hand side vanishes and one obtains the correct expression for GR. In the presence of the higher derivatives the leading term of the function $F(z)$ at small $z$ is
\be\n{Fexp}
F(z)\sim C-{1\over 4}S_2 z(\ln z-2c)-{1\over 4} S z+O(z^2)\, ,
\ee
where $c=1+\ln 2-\gamma$ and $S=\sum_{i=1}^n \mu_i^2 \ln(\mu_i^2) P_i^{-1}$. For the ghost-free gravity one has \cite{FZT}
\be
F(z)=\ln z +\gamma +\mbox{Ei}(1,z)\sim z-{1\over 4}z^2+O(z^3)\, .
\ee

The obtained metric (\ref{METR}), (\ref{PPP}) can be used to find a solution for the linearized HD and GF gravity equations for a collapsing spherical thin null shell. For this purpose one considers a set of "photons", passing through a fixed point $P$ of the Minkowski spacetime. In the continuous limit this set fills the surface of the null cone, with the vertex at $P$. We additionally assume that the density of this spherical distribution of the "photons" is uniform and the corresponding mass per a unit solid angle is $M/4\pi$.   Since we are working in the linear approximation the resulting gravitational field for such a distribution is
$ds^2=ds_0^2+\lan dh^2\ran$, where $\lan dh^2\ran$ is obtained by averaging of a single "photon" metric over their spherical distribution. The calculations give \cite{FZT}
\ba
&&ds^2=-dt^2+dr^2 +r^2 d\omega^2+\lan dh^2\ran\, ,\ \ z=r^2-t^2\, ,\nonumber\\
&&\lan dh^2\ran=-2GM r^{-1} F(z)[(dt-t dr/r)^2+ z d\omega^2/2]\, .\n{AVER}
\ea
Let us denote
\be
g=(\nabla\rho)^2\hhh   \rho^2\equiv g_{\theta\theta}=r^2-{GM\over r} z F(z)\, ,\\
\ee
then the equation $g=0$ determines a position of the apparent horizon, if the latter exists. In the linear in $M$ approximation this function is
\be\n{ggg}
g=1-2G M  r^{-1}q(z)\hh q(z)= z F'(z)\, ,
\ee
where $(\ldots)'=d(\ldots)/dz$. For GR (as well as for GB and $L(R)$ gravity) $q(z)=1$.

Using (\ref{FHDG}) one finds that for the HD gravity
\be
q(z)=1-\sqrt{z}\sum_{i=1}^n \mu_i P_i^{-1} K_1(\mu_i \sqrt{z})\, .
\ee
For small $z$ one has
\be\n{qez}
q(z)=-{1\over 4} S_2 z(\ln z -2c+1)-{1\over 4}S z+O(z^2)\, .
\ee

Let us demonstrate now that  the function $g$ is positive for small enough $M$, and, hence, the apparent horizon does not exist. Let us notice that outside the null shell $|t|/r<1$. We denote $t=\pm\sqrt{1-\beta^2}$, $0\le \beta\le 1$, then one has ($y_i=\beta \mu_i r$)
\be
q(z)/ r=\beta \sum_{i=1}^n \mu_i P_i^{-1}Z(y_i)\hhh Z(y)={1\over y}-K_1(y)\, .
\ee
The function $Z(y)$ is positive and takes maximal value $0.399$ at $y=1.114$. Thus
\be
|q(z)|/ r< 0.4 \sum_{i=1}^n \mu_i |P_i|^{-1}\, .
\ee
This implies that for small enough value of the mass $M$ the invariant $g$ is positive everywhere outside the shell. In other words, for such mass $M$ the collapse of the null shell does not produce a mini black hole. This means that for the class of the higher derivative theory of gravity (\ref{Qnn}) with $n\ge 1$ there is a mass gap for the mini-black hole production. The value of this map is determined by the characteristic length scale $\mu^{-1}$ of the theory. The apparent horizon does not exist if $M\mu\lesssim 1$. The same conclusion is valid for the GF theory of gravity \cite{FZT}.

It is possible to calculate the curvature invariants for the metric (\ref{AVER}). In particular, the  Kretschmann curvature invariant ${\CAL R}^2$ in the lowest order in $M$ is
\be
{\CAL R}^2={48 G^2 M^2\over r^6}{\CAL F}\hh
{\CAL F}=2z^2 {q'}^2-2z q q' +q^2\, .
\ee
Using expression (\ref{qez}) for small $z$ one finds
\be
{\CAL F}\sim {1\over 16} z^2[(w^2+4w+5)S_2^2 +2(w+2) S S_2+S^2]\, ,
\ee
where $w=\ln z-2c$. This means that Kretschmann curvature vanishes on the null shells. However, in a general case it is divergent at $r=0$.

We demonstrate now that this divergence is a result of the (unphysical) assumption, that the thickness of the null shell is zero, and for the collapse of the shell with the finite thickness this divergence becomes softer or is absent. To obtain a solution of the HD gravity equations for such a thick shell we proceed as follows \cite{FZT}. Consider a set of spherical null shells collapsing to the same spatial point $r=0$, but passing it at different moments of time $t$. In the continuous limit, one obtains a distribution of the matter, that describes a spherical thick null shell which initially collapses and has a mass profile $M(t+r)$, and after passing through the center it re-collapses with the mass profile $M(t-r)$. In the linear in $M$ approximation the gravitational field of such a shell can be obtained by averaging the solutions (\ref{AVER}). We denote by $\llan dh^2\rran$ the result of the averaging of the perturbation $\lan dh^2\ran$. For simplicity we present here the expression for $\llan dh^2\rran$ for the case when $\dot{M}$ is constant, and the time duration of the thick shell is $b$, so that the total mass $M$ of the shell is $\dot{M} b$. In the domain of the intersection of the incoming and outgoing null fluid fluxes the metric is static. The calculations give (see \cite{FZT} for more details)
\be
\llan dh^2\rran=-{2GM\over br}[c_0 dt^2+c_2 {dr^2\over r^2}+
{1\over 2}(c_0 r^2-c_2) d\omega^2]\, ,
\ee
where $c_k=\int_{-r}^r dx x^k F(r^2-x^2)$. It is easy to check that constant $C$, which enters (\ref{Fexp}) does not contribute to the curvature. For this reason we put $C=0$.
Using expansion (\ref{Fexp}) of $F(z)$ one obtains
\ba
&&c_0=-{r^3\over 9}[(6u-5)S_2+3S]\, ,\nonumber\\
&&c_2=-{r^5\over 225}[(30u-31)S_2+15S]\, ,
\ea
where $u=\ln r-c-\ln 2$. For small $M$ the function $g$ remains positive, while the Kretschmann invariant is
\be
{\CAL R}^2\sim{32\over 27} G^2\dot{M}^2[(36 u^2+5) S_2^2 +36 u S_2 S+9 S^2]\, .
\ee
Hence,  a collapse of a thick null shell in the theory with higher derivatives results in the logarithmic singularity of the curvature. However, if such a theory is $\psi$-regular, the curvature is finite. In particular, this property is valid for any regular theory with higher derivatives.
For the ghost-free theory of gravity the Kretschmann invariant ${\CAL R}^2\sim {32\over 3} G^2\dot{M}^2\mu^4$ is always finite at $r=0$ \cite{FZT}.

Let us summarize the results. Our main conclusion is that in the higher dimensional and ghost free theories of gravity there exist a mass gap for a mini black hole production. Such theories possess a characteristic length scale $\mu^{-1}$ and the apparent horizon is not formed when the gravitational radius of the mass $M$, $2GM$, is smaller that $\mu^{-1}$. For a regular (and $\psi$-regular) theory the Kretschmann curvature invariant for the thick null shell collapse is regular. Whether in the strong gravity limit this property remains valid, and the black hole interior does not contain singularities, is an interesting open question.

The author thanks the Killam Trust and the Natural Sciences and Engineering Research Council of Canada for their financial support.

\end{document}